# Guiding waves along an infinitesimal line between impedance surfaces

Dia'aaldin J. Bisharat[1,2*] and Daniel F. Sievenpiper[2*]

[1]*Department of Electronic Engineering, City University of Hong Kong, Kowloon, Hong Kong, China*
[2]*Electrical and Computer Engineering Department, University of California, San Diego, California, 92093, USA*
*\*dbisharat2-c@my.cityu.edu.hk; dsievenpiper@eng.ucsd.edu*

**We present a new electromagnetic mode that forms at the interface between two planar surfaces laid side by side in free space, effectively guiding energy along an infinitesimal, one-dimensional line. It is shown that this mode occurs when the boundaries have complementary surface impedances, and it is possible to control the mode confinement by altering their values correspondingly. The mode exhibits singular field enhancement, broad bandwidth, direction-dependent polarization, and robustness to certain defects. As a proof-of-concept, experimental results in the microwave regime are provided using patterned conducting sheets. Our proposed effective-medium-based approach is general, however, thus allowing for potential implementation up to optical frequencies. Our system is promising for applications including integrated photonics, sensing, switching, chiral quantum coupling, and reconfigurable waveguides.**

Having a peak field bound to the interface of two media makes surface waves (SWs) [1] attractive for signal transmission and routing with simple implementation for sensing and communication applications [2,3]. Exploiting surface plasmon polaritons' (SPPs) nature [4], SWs can also exhibit strong light confinement which is useful for realizing subwavelength guiding structures and as a result potentially high-density integration of optical circuits and lower waveguide bending loss. Variations such as V-shaped grooves [5] and metallic wedges [6] have been proposed, albeit with more complexity, which reduce SWs to one dimension despite the absence of an enclosing structure, thus enabling greater guiding control. Similarly, guided modes at the edge of photonic crystals (PCs), within limited simultaneous bulk and surface bandgaps, have been proposed [7]. In addition, edge plasmons have been observed in graphene ribbons [8].

Recently, there has been a special attention to photonic structures with interface modes exhibiting robust directional propagation. Notably, this includes photonic topological insulators (PTIs) based on symmetry-protected topological (SPT) phases [9,10,11,12]. In these systems, where time-reversal (TR) symmetry is not broken [13,14,15,16], crystalline or intrinsic symmetries of the wave fields and differing topology of bulk bands give rise to wavevector-locked states at the interface [12]. Analogously, opposite single-negative bulk materials [17] support bound states that exhibit a similar though limited robustness [18,19,20]. In addition, trivial structures such as nanofibers and glide-plane PC waveguides [21], where light is tightly confined with evanescent wave on their interface, can exhibit direction-dependent polarizations [22].

In this work, we introduce a new one-dimensional mode, analogous yet different from SWs, that is confined to the line interface between two planes and propagates in air without the need for any enclosing structure. Besides forming the smallest waveguide possible, the line wave has robust wavevector-locked states, broad bandwidth, and does not require any bulk media. In addition, it exhibits strong -ideally singular- field enhancement and tunable spatial field confinement. We establish conditions for the mode's existence by characterizing the interfaced planes merely by complementary isotropic surface impedances. Then, we provide measurement results proving the feasibility of using periodic surfaces with certain effective medium properties, such as simple frequency-selective-surfaces (FSS). Furthermore, we examine the characteristics of our system at different conditions and suggest possible applications. Our work paves the road for planar, compact and efficient routing and concentration of electromagnetic energy in the microwave through optical regime and opens the door for unconventional devices.

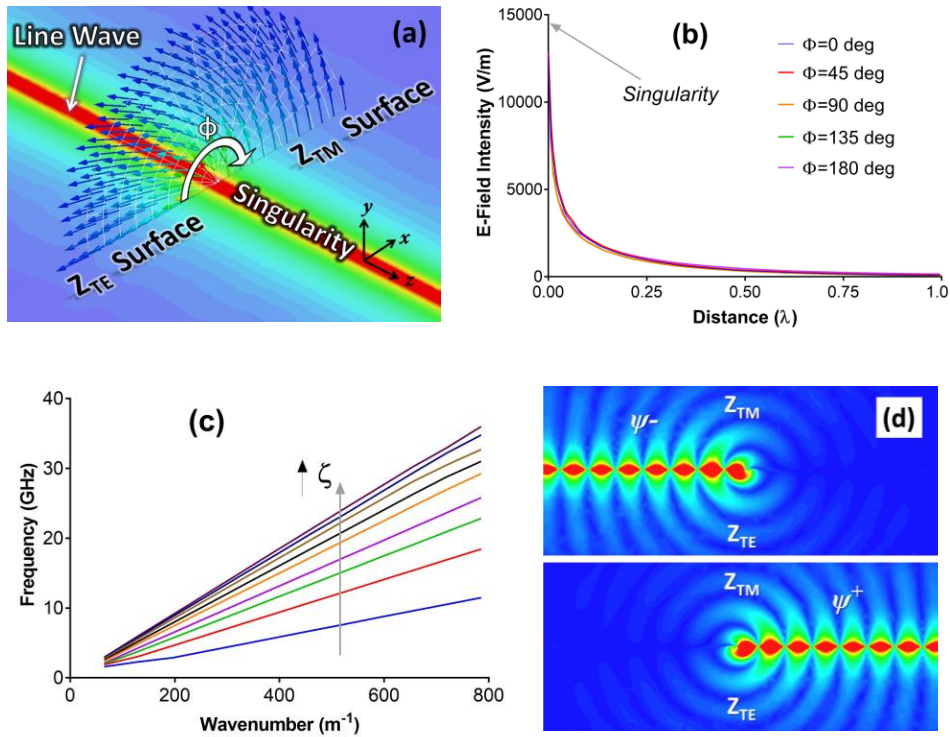

FIG. 1 (color online). Simulated fields characteristics of the line wave mode. (a) Magnitude distribution and vector plot of the electric field above the TE and TM surfaces across the line interface showing the line wave mode (linear scale), (b) decay curves of the electric field at different directions around the line interface, (c) dispersion curves of the line mode at different values of complementary impedances across the interface, and (d) pseudo-spin states excited by electric and magnetic Hertzian dipoles along y-axis in phase (above) or out of phase (below) at the interface.

SW modes can be guided on planar structures with subwavelength periodic inclusions, known as metasurfaces [23,24], whose response to impinging waves and their guiding properties can be conveniently characterized by surface impedance [25,26]. This methodology has been extensively used for variety of applications including EM guiding, absorption, radiation, scattering alteration, cloaking, and self-focusing [27,28,29,30,31,32]. Consider a field with an exponential decay $e^{-\alpha y}$ away from the surface and a propagation function $e^{-j\beta z}$, such that $\alpha^2 = \beta^2 - k^2$, where $k$ is the wavenumber in the free space. The surface impedances for transverse-magnetic (TM) and transverse-electric (TE) polarized waves are [33]:

$$Z_{TM} = j\eta_0 \frac{\alpha}{k} \quad , \quad Z_{TE} = -j\eta_0 \frac{k}{\alpha} \tag{1}$$

where $\eta_0$ is the intrinsic impedance of free space. Meanwhile, the refractive index $n$ seen by the SW is $n = c/v_p = \beta/k$, where $c$ is the speed of light in free space and $v_p$ is the phase velocity of the wave along the surface. Therefore, the relationship between the surface impedance and the refractive index is given by [34]:

$$Z_{TM} = \eta_0 \sqrt{1-n^2} \quad , \quad Z_{TE} = \eta_0 / \sqrt{1-n^2} \tag{2}$$

Accordingly, for TM- and TE-polarized SWs propagating at equal phase velocities, we can define a new parameter $\zeta$ that relates the two respective surface impedances as follows:

$$j\frac{Z_{TE}}{\eta_0} = j\frac{\eta_0}{Z_{TM}} = \zeta \tag{3}$$

First, consider the case where $\zeta$ is infinite, so that the TM-surface becomes a perfect electric conductor (PEC) while the TE-surface becomes a perfect magnetic conductor (PMC). Obviously, the PEC (PMC) surface forces the tangential electric (magnetic) field to vanish, thus allowing only the TM (TE) SW mode to survive. When interfacing PEC and PMC surfaces, a new mode that is localized at the interface appears. The new mode is a product of interference between TM and TE modes, which are only supported at either side of the interface. As a result, given the open boundary nature of the resulting structure, the mode decays away from the interface line. As the cross-sectional view in Fig. 1a shows, the electric field vectors point in the transverse direction adjacent to the $Z_{TE}$ surface and vary gradually toward the normal direction as we trace a path at constant distance away from the interface line towards the $Z_{TM}$ surface. Hence, it is straightforward to represent this waveguide configuration in cylindrical coordinates and the waveform of the mode is deduced as [35] (see Supplementary Material [36]):

$$E_z = E_0 K_{\frac{1}{2}}(\alpha\rho) \operatorname{Sin}\left(\frac{\phi}{2}\right) e^{-j\beta z} \quad , \quad H_z = \frac{E_0}{\eta_0} K_{\frac{1}{2}}(\alpha\rho) \operatorname{Cos}\left(\frac{\phi}{2}\right) e^{-j\beta z} \tag{4}$$

where $K$ is the modified Bessel function of the second kind, and $\alpha^2 = k^2 - \beta^2$, with $\beta \geq k$.

The waveform is verified with full wave simulation in ANSYS HFSS software, which clearly shows the singular nature of field intensity at the interface line as depicted in Fig. 1b. Here, for a given $\zeta$ value, the field intensity decays away from the interface at different $\phi$ angles at the same rate. Note that although the field is infinite at the line, the field everywhere has a finite integral and thus the power carried by the line wave is finite. Just as SWs on good conductors are only loosely bound to the surface, this is also the case for line waves, which have $\beta = k$ for the limit of a PEC-PMC interface. A more tightly bound mode is readily attainable by adopting a finite $\zeta$ value, hence $\beta > k$, as shown in Fig. 1c, with smaller $\zeta$ resulting in slower mode. Note that regardless of $\zeta$ value, the field remains infinite at the line in the absence of loss (see Supplementary Material [36]).

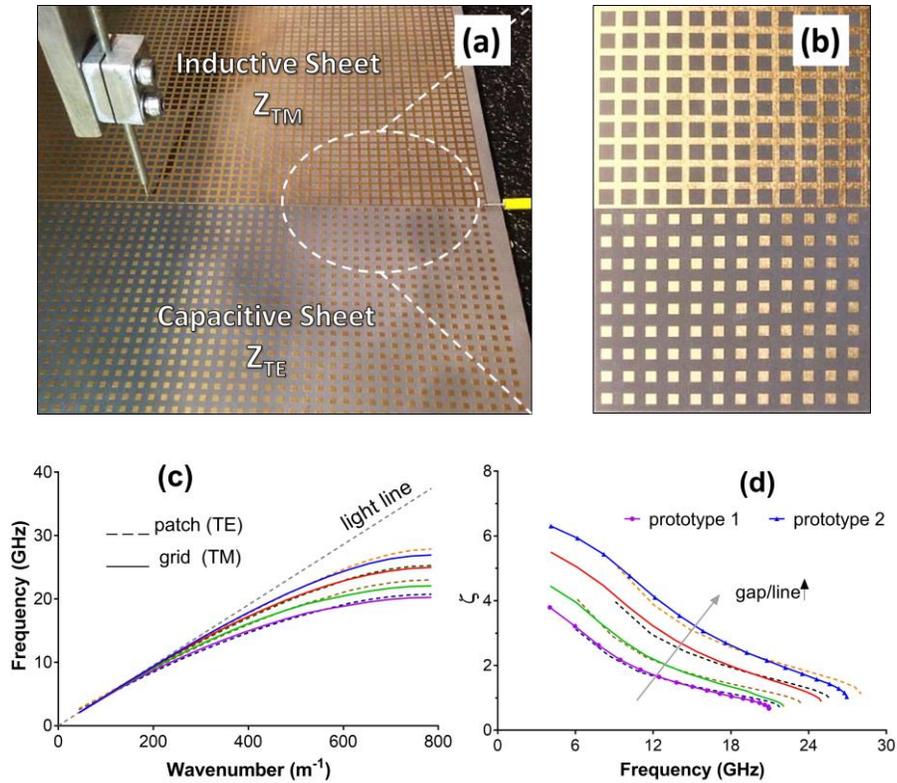

FIG. 2 (color online). Measurement setup and characteristics of the fabricated FSS sheets. (a) A probe antenna (right) oriented along the interface line is used as the excitation source while another probe (left) oriented vertically at a 2mm distance above to the surface is used to scan the relative intensity of the normal component of electric-field, (b) zoom-in of the complementary FSS sheets fabricated on a printed-circuit-board (PCB) on Rogers 5880 ($\varepsilon_r = 2.2$, $\delta t = 0.001$) substrate with a 0.8mm thickness, (c) dispersion characteristics of TM and TE FSS cells of different sizes, and (d) $\zeta$ values versus frequency for different sizes of FSS cells. Prototypes 1 and 2 both have unit cell period of 4mm. Prototype 1 (2) has grid line width of 0.2mm (1.2mm) and gap width of 0.8mm (2.2mm) between patches. The difference between the sizes of the complementary FSS cells is to compensate for the substrate permittivity.

To realize the line wave, two surfaces whose impedances take the form in equation (3) are required, that is an inductive surface to support a TM mode and a capacitive surface to support a TE mode. These criteria can be fulfilled by simple frequency-selective surfaces (FSS) such as these shown in Fig. 2b. Here, the conducting patches (grid) exhibit a dominant capacitive (inductive) response at frequencies where the FSS cell is subwavelength. Fig. 2c shows that the respective SW modes of the complementary surface have dispersion curves that overlap over a wide frequency band. Therefore, TM- and TE-mode of the same phase velocity [37] are supported and a line wave could be supported at their interface. Two prototypes with interfaced FSS sheets were prepared, whose constitutive unit cells' size corresponds to about $\lambda_0/12$ at 6GHz and to about $\lambda_0/4$ at 18GHz, where $\lambda_0$ is the wavelength in free space. Fig. 2d shows the associated $\zeta$ values at different frequencies, which are higher in the case of prototype 2.

Fig. 3a, which maps the relative intensity of the normal component of electric field measured at a fixed distance above the impedance surfaces, shows successful excitation and transmission of the line wave along the interface for a distance of several wavelengths at different frequencies. Fig. 3b shows the spatial field confinement across the interface line obtained from experiment as well as simulation under realistic geometry and material parameters. The electric field component normal to the surface is evident on both sides of the interface, as expected, albeit with slightly larger amplitude on the TM side (the positive x-axis of Fig. 3b). Moreover, due to lower $\zeta$ values, prototype 1 exhibits greater field concentration than prototype 2, as expected. The simulated and measured results across the interface at roughly 1 ~ 2mm (limited by our near-field measurement setup) above the surface are in good agreement. On the other hand, the simulated field intensity at the center of the line interface shows higher enhancement level (see Supplementary Material [36]) with an effective mode width (half-power concentration) of less than $\lambda_0/15$.

The measured operation range, which spans about two octaves of bandwidth, could be extended by adopting other complementary artificial surfaces with lower dispersion and whose dispersion curves overlap over a wider frequency range. In addition, the fields at the singularity are limited in physical implementations by the thickness of the surface, the dissipation losses, and the periodicity of the FSS structures (see Supplementary Material [36]). At higher frequencies up to the optical domain, it is thus more suitable to use 2D materials (i.e., truly homogeneous surfaces) whose properties can satisfy the operation criteria given in equation (3) at the frequency of interest. One potential example of such materials is graphene at terahertz regime, which is known for its highly-confined long-lifetime plasmons [38]. Specifically, graphene can be modeled as an impedance sheet and support TM and TE surface modes depending on its doping level [39]; hence allowing straightforward implementation and tunabilty thanks to the scalability and universality of the proposed effective surface impedance approach.

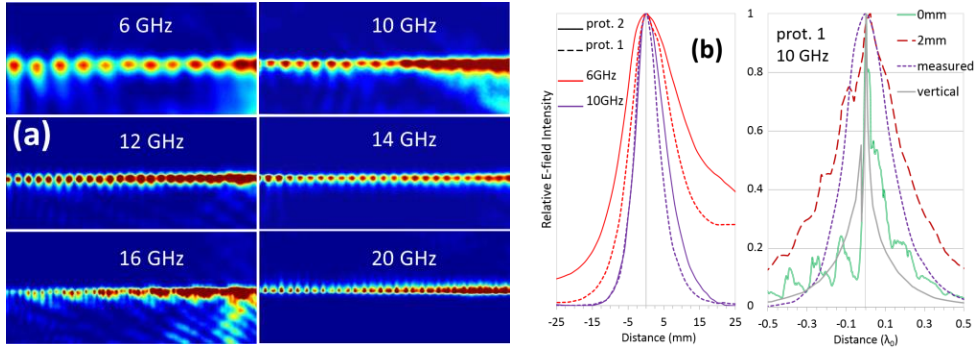

FIG. 3 (color online). Measured results of the line wave. (a) Electric field magnitude distribution at different frequencies on top of prototype 1 (left) and prototype 2 (right), and (b) normalized decay curves of the normal component of electric field in the transverse direction to the line interface (linear scale) at a fixed finite distance above the two prototypes (left) and comparison with simulated results at a distance of 2mm and 0mm from the interface (right). Simulated fields through origin are also plotted along the normal direction to the surface, which show no ripples except for a dip in field intensity residing inside the substrate.

Another important aspect is that the line wave exhibits wavevector-locked states. Unlike in a medium like air, PEC and PMC boundaries do not give rise to two decoupled states naturally since they have opposite effects on the electric and magnetic field components. However, joined PEC ($\varepsilon = -\infty, \mu = 1$) and PMC ($\varepsilon = 1, \mu = -\infty$) boundary conditions may preserve electromagnetic duality, for instance by forming a pair of mirror images about the *yz* plane satisfying the $\varepsilon(x) = \mu(-x)$ inversion-symmetry [40,41]. Note that a single PEC-PMC interface is sufficient to partially bound energy due to $\varepsilon$-negative and $\mu$-negative materials possessing different topological orders when considering a fixed wave polarization [18,19,20]. Clearly, both materials support evanescent waves but possess opposite signs of imaginary impedance. As a result, new modes emerge that are a combination of magnetic and electric modes with a specific phase relationship, hence conserved pseudo-spin values (see Supplementary Material [36]) [42]. Furthermore, since the pseudo-spin configuration is uniquely defined by the direction of the propagation wavevector ($\beta$), the interface constitutes a spin-filtered channel [10,40]. This makes our system somewhat reminiscent of time-reversal (TR) invariant symmetry-protected topological (SPT) states formed between two claddings of opposite bianisotropy [12], where intrinsic symmetries of the wave fields and differing topology of bulk bands give rise to counter propagating pseudo-spin states up ($\psi^+$) and down ($\psi^-$) (see Supplementary Material [36]).

The approach above can be generalized to include the formation of domain walls by the inversion of surface impedance; that is interfacing inductive/capacitive surfaces with identical $\zeta$ value. Note that besides offering design flexibility and additional control over the properties of the interface states, this generalization solves the issue of the weak cross coupling between TM and TE modes in the PEC-PMC case, which otherwise necessities using a closed waveguide configuration for practical applications. The paradigm of effective surface impedance has been exploited in relation with band geometric (Zak) phases to explain the appearance of interface states in 1D and 2D systems of PCs [43,44]. In comparison, our system is free of the bandwidth limitation associated with bandgap in PCs and supports direction-dependent polarizations as evident in the full-wave simulation shown in Fig. 1d. Note that although the spin-momentum locking property is universal in evanescent waves [22], it is more prominent in the case of the line wave due to the strict confinement in the transverse plane to the wavevector leading to 1D propagation only.

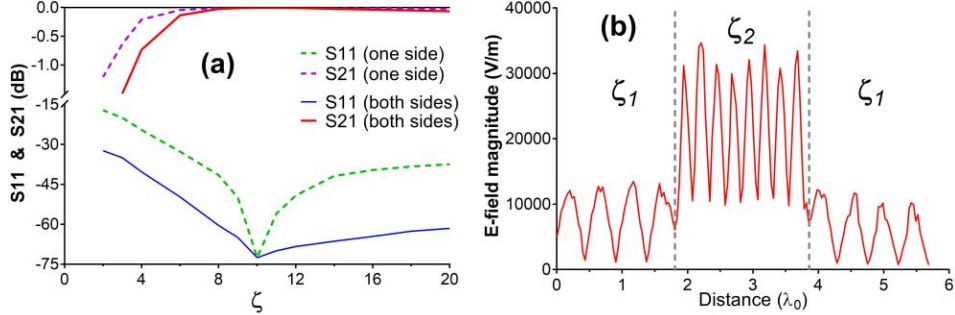

FIG. 4 (color online). Influence of a finite discontinuity in surface impedance on the line wave. (a) Transmission and reflection coefficients of the mode due to a defect ($\zeta \neq 10$) at one or both sides of the interface over a distance of 0.6 $\lambda_0$, and (b) snapshot of field magnitude depicting an increased enhancement and a shorter wavelength over a finite distance due to change in surface impedances across the line interface.

The spin-momentum locking feature enforced by the boundary inversion symmetry endows our line guide with robustness against reflection from certain structural defects. To qualify this symmetry-protection, we introduce a discrete discontinuity in surface impedance over a finite distance along the interface line (see Supplementary Material [36]). As shown in Fig. 4a, for a large impedance variation ($6 \leq \zeta \leq 20$) in either one or both impedance surfaces, reflection coefficient ($S_{11}$) lower than -30dB and isolation ($S_{21}$) of about -0.1dB are achieved. This is expected given that such defect does not violate spin-degeneracy or cause reversal of boundary conditions. The larger reflection in case of one-sided impedance discontinuity is because of the degraded boundary symmetry, which leads to mismatch in phase velocity across the interface. Note that the wave may also couple into free space due to the open-boundary nature of the system. This result, though does not show complete immunity to backscattering, is to some extent similar to topological protection in [12] which exists as long as a bandgap separating two inverted bands does not close (see Supplementary Material [36]).

By taking advantage of this robustness, we could intentionally alter the surface impedances (i.e. $\zeta$ value), simultaneously across the interface to control the spatial field confinement of the line mode and its propagation constant as shown in Fig. 4b. For example, this enables the design of compact delay lines and phase shifters without the need for any bends and without occupying additional space. On the other hand, switching the sign of the surface impedance across the interface would forbid propagation of the interface mode. This feature is useful for building network devices with simple implementation in microwave and photonic applications. For example, simulated Fig. 5a shows that the junction due to the surface impedance reversal forms a four-port network that functions as a common magic-T structure, where the line wave fed at port 1 is guided to ports 2 and 4 with no energy coupling to port 3 as desired.

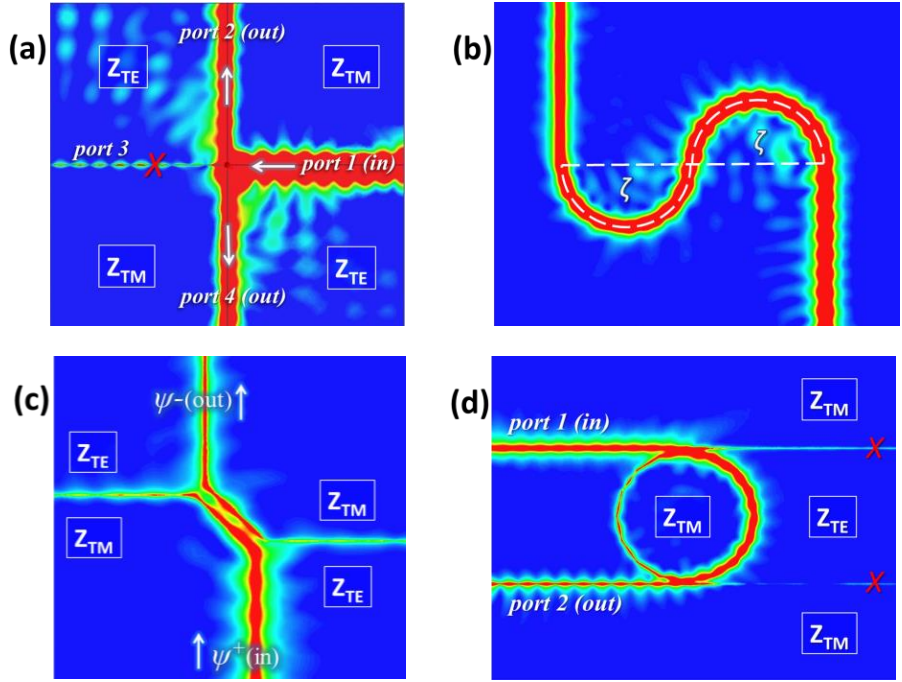

FIG. 5 (color online). Full-wave simulations for potential applications of the line wave. (a) Wave transmission in four-port network (magic-T) of a junction due to surface impedance reversal, (b) wave transport along a curved interface line with a proper change in TM and TE surface impedances restoring equal phase velocity at the two sides, (c) coupler structure showing excitation of a reversed pseudo-spin mode, and (d) an implementation of a ring-resonator showing frequency selection.

One the other hand, guiding the line mode along a bent path seems to cause scattering (see Supplementary Material [36]). This is due to the curvature causing the phase velocities of the wave on both sides of the interface to differ, thus resulting in a leakage similar to that of conventional SW mode along a curved surface. This is different from PTIs and PC defect waveguides, which have a complete bandgap in the bulk where no modes can exist. Although the eigenfields of the proposed line mode must be localized near the interface, energy can couple to the SW modes that are supported at the same frequency on the surrounding surfaces. Nevertheless, we can prevent such leakage by compensating for the phase velocity difference through adjusting the relative impedance values; hence restoring the original relative boundary conditions, as depicted in simulated Fig. 5b. Here, a relatively lower $\zeta$ value is used in the dotted semi-circular area inward of the curve in order to increase the propagation constant of the SW mode in that region so that the two phase fronts across the interface would move along the line at the same speed. This ability to couple to SW modes can be useful for unique applications. For example, Fig. 5c shows a coupler design that enables transferring energy between two eigenfields of opposite pseudo spin-polarizations in nearby waveguides. In addition, the line wave can be used in a ring resonator design as illustrated in Fig. 5d for filtering applications.

The above findings make the line wave appealing for energy confinement and transport as well as integrated photonics applications, as a one-dimensional object being the smallest waveguide possible. In addition, due to its planar configuration, strong mode confinement [45] and pseudospin-polarization [42], the line wave is useful for light-matter interaction and chiral quantum processes (see Supplementary Material [36]), where the interaction depends on the light's propagation direction and the emitter's transition dipole moment polarization [21]. Moreover, the significant field enhancement of the mode as well as having an air channel at the line interface can be potentially used to free electrons from materials and guide them in air free of scattering, thus offering simple implementation of high speed micro-plasma and vacuum based electronic devices [46]. Furthermore, our approach based on surface impedance boundary conditions allows for line waves with reconfigurable pathways [47], for instance, through electrostatic biasing in graphene [39]. This tuning capability in addition to the field singularity can also pave the way to nonlinear photonic structures for switching and modulation applications [48].

This work has been supported in part by AFOSR grant FA9550-16-1-0093.